# The Marginalized Identities of Sense-makers: Reframing Engineering Student Retention

Brian A. Danielak, Ayush Gupta, Andrew Elby
University of Maryland, College Park, briandk@umd.edu, ayush@umd.edu, elby@umd.edu

*Abstract* – This paper empirically argues for a closer examination of what we wish to retain when we speak of "retention" in engineering [1]. We present and interpret data from clinical interviews and classroom video of "Michael," a student who feels marginalized by an engineering program that undervalues him because of his stance toward knowledge [2],[3]. Michael is a sophomore Electrical Engineering and Mathematics major in a Basic Circuits course. In his own words, he's a "fringe" student because of his robust tendency to try making sense of the concepts being taught rather than memorizing formulae. He also feels alienated because he views learning in terms of argument and intuition, not algorithm and rote acceptance. Furthermore, for Michael the practice of sense-making defines him; it's an integral aspect of his identity [4]. Thus, Michael's self-reported sense of alienation resonates strongly with existing identity-based accounts of students leaving the field [5],[6]. We contend the field of engineering suffers if individuals like Michael don't pursue it. Through this case study of Michael, we urge the retention discussion to consider not just the demographic categories of people we hope to keep, but also the approaches to knowledge, learning, and problem-solving we aim to support.

INTRODUCTION – "AS AN INDIVIDUAL, OR…?"

Michael caught our eye as one of the most outspoken students in his Spring 2009 Basic Circuits class. We videotaped and analyzed discussion sections of his course, and Michael could consistently be seen up at the board, discussing the meanings of equations and graphs, or debating the merits of a particular approach to analyzing the circuit. At the end of the semester, the first author began an ongoing series of one-hour semi-structured interviews with Michael. Less than six minutes into our first interview—in which Michael speaks of the conceptual meaning of first-order differential equations for circuits—we encountered the question that motivates our paper.

"With first-order," Michael says, "you can look at it and say 'I know what's gonna happen' even before you do the calculations. So, when you get an answer you can tell whether it makes sense or not."

"How do you know?"

"You just get a feel for how things like capacitors and inductors behave in the long run. If a capacitor should have an open circuit potential, then a voltage of zero as *t* approaches infinity just doesn't make sense. Inductors and current work in a similar way."

"As you go through a problem, is that something you think about? Do you ask yourself whether your mathematical answer makes sense with what you know about the circuit should behave?"

The interviewer can't finish the question; Michael cuts him off.

"Are you asking me as an individual," he says, smiling, "or as a representative of the people in my class?"

SENSE-MAKING AND RETENTION IN THE LITERATURE

Michael's story touches on two distinct themes of STEM education research. First, in interviews Michael speaks tirelessly of mathematics "making sense" with what he knows about the physical world. The vignette above is just one example of a consistent pattern. Michael insists that to him, mathematical equations do not simply compute; rather, they *embody* persuasive, intuitive accounts of how the world behaves [7-9]. As a result, Michael's remarks about sense-making are *epistemological*: they connect the way he uses math to his views of what it is to *know* and *understand* in mathematics and engineering [10],[2],[3].

The second theme is less obvious, but no less pertinent. As we show below, Michael thinks his views about mathematics set him apart from most other engineering students. His epistemological views also contribute to his growing sense of contrast between how he thinks engineering courses *should* be taught and assessed and his experiences of how they *are* taught and assessed. Such issues fall squarely in the domain of engineering student retention: discussions of how to keep students in engineering [1],[11],[12], how to reshape the plurality of engineering [13-15], and the often-complex stories of why some students leave the field [5],[6],[16].

*I. Sense-making as Mathematical Practice*

Educational research argues for the importance, and indeed primacy, of mathematical and physical sense-making in the physical sciences and engineering [7],[17],[8],[18],[9],[10]. As one physics education researcher put it, "[w]e do not want meaningless symbol manipulation; if students use

In Review

symbolic expressions, we want them to use the symbols with understanding" [7].

One way to understand sense-making is as an ongoing interplay between *internal coherence*—the agreement of formalisms with one another—and *external coherence*—the alignment of formal statements with what we know of the outside world. For example, as Schoenfeld [9] discusses, external coherence has broken down when elementary school students who correctly performed a long division calculation chose "31 remainder 12" as the number of buses needed to move a group of soldiers [9]. Sherin [7] shows how college physics students' sense-making is supported by their use of cognitive structures in which a mathematical symbol template and associated conceptual content are compiled into a single unit of thought [7]. In Sherin's framework, internal and external coherence are strongly interconnected, both within the conceptual content of physical expressions and their formal representations as mathematical symbols [7].

Studies of professional engineering practice {refs}, in which sense-making is ubiquitous, suggest that engineering education should support a student's search for coordination among representations (internal coherence) and their relation to the world (external coherence) [8],[19]. In analyzing Michael, we thus attend to his attempts to find internal and external coherence and also his views about how those searches for coherence relate to professional engineering practices.

*II. Retaining Students in Engineering*

There are longstanding concerns about how to keep students in science and engineering [16],[13] that continue to connect to engineering education research [1],[11]. The "leaky pipeline" model has been especially useful as an analytic metaphor [20],[14],[21],[15]. But, that model has come under fire more recently by research that pushes for a finer-grained reconception of retention [5],[6]. Stevens et al argue from an ethnographic perspective that the mechanical pipeline analogy obscures the very complexity of the students we seek to retain [6]. Michael illustrates this point: Understanding his sense of alienation as an engineering student requires a fine-grained analysis of how his epistemological views about the nature and importance of sense-making color his day-to-day experiences in his classes.

### THEORETICAL FRAMEWORK AND METHODS

Studies of marginalization and identity benefit from qualitative research approaches that are sensitive to the challenges of depicting subjects in an authentic, trustworthy, socially responsible way [22],[23]. We further sought to "recover the person" in our research narrative, both in individual complexity and discursive importance [24], by observing him in both discussion section and lectures, interviewing him about those and other experiences, and allowing Michael to drive the direction of many of the interviews.

Our primary data came from both in-class observation and semi-structured clinical interviews. We reviewed one-hour videotapes of each of Michael's weekly discussion sections in his Circuits course, and the first author has thus far conducted five one-hour semi-structured interviews with him. The interviews typically contained a mix of interactive prompts, in which Michael solves a problem from his homework or addresses a new problem we posed, and reflective prompts, in which Michael talks about his experiences inside and outside school. These prompts often arose from spontaneous statements Michael made, such as the distinction he hinted at between his own sense-making practices and those of his classmates (from the Introduction to the paper). With each interview, the research design evolved [22] to pursue emergent questions of both Michael's sense-making practices and their connection to his identity [4]. Throughout, working as a group, we proposed and debated interpretations of classroom activity and interview utterances, gestures, and expressions, seeking confirmatory and disconfirmatory evidence in the data [25].

In coordinating across contexts—from classroom to interview, and from 2009 to 2010—we strove not just to support emerging hypotheses through triangulation, but also to achieve Richardson's idea of methodological crystallization: a "deepened, complex, thoroughly partial understanding" of Michael on his own terms [26].

### EMPIRICAL ARGUMENT OVERVIEW

In the following sections, we present and analyze selections from several one-hour clinical interviews we conducted with Michael. We also interpret portions of Michael's activity in a basic circuits discussion section. We argue two intertwined points, mutually supported by and triangulated around our concurrent analysis of in-class and in-interview data:

*(1)* Michael is aware that his epistemological stance toward sense-making is different from—and at times opposed to—the stance his classmates take. He sees himself as proceeding from his own epistemological convictions, which he has come to see as extensions of his own identity that are rooted in practice. But, he views the field and his school program as at best indifferent to, and at worst biased against, his personal convictions about the importance of sense-making in engineering.

*(2)* Michael positions himself differently from others in his discussion section because of his distinct epistemological stance. In short, many students are looking for the "right answer" while Michael is looking for an answer that is both physically and mathematically convincing. In practice, Michael seeks not just evidence but explanation. He appeals to different warrants for what "counts" in arguments about how circuits behave.

### SENSE-MAKING IS PART OF MICHAEL'S IDENTITY

*I. I would much rather have an essay test*

Michael sees the practice of sense-making as marking him apart from "the people in my class." In the vignette we started to present in the Introduction, he takes pains to distinguish himself from others:

> Michael: You're interested in Michael? OK. So. Yes, I always do that to see if it makes sense. I always—in fact I don't even really like doing the formulas to begin with. I hate the computation aspect of the class. I would much rather have it be an essay test and be able to talk about everything that's goin' on. (May 13, 2009)

Michael is aware that his stances toward learning, knowing, and conceptual reasoning seem out of the mainstream:

> Michael: Some people say it's a good thing, some people say it's an illness, but I have a lot of pride [in my sense-making]…. I feel that if you say something [on a test] that makes absolutely no sense, like that's just the worst thing for me.

Michael doesn't think his views about learning are supported by the structure and coverage of typical engineering courses. On a circuits exam, for example, he did particularly well on conceptual questions that required reasoning and argumentation. But, those questions were designated "extra credit," signaling that they test the periphery rather than the core of what students should know. For Michael, the "extra credit" conceptual reasoning parts *are* the core. He's said so plainly, across several interviews:

> Michael: What [that exam] was intended for was, so, if people got none of the conceptual things right they could still do well. But, it sorta was the opposite for me. The extra {air quotes} "credit" helped balance out for the things that I should have been able to just regurgitate. (May 13, 2009)

Michael's views go beyond the role of extra credit. Since we began interviewing him in May of 2009, his position on what counts as evidence of knowledge and learning have been remarkably consistent. He feels "the current education system we have [in engineering] doesn't reward good learning so much as it rewards regurgitation, and good memory" (March 17, 2010). At stake for Michael are long-standing issues about the kinds of knowledge engineers are responsible for [6], and what constitute "real" problems in the discipline [27].

> Michael: What I'd like to see, if it's possible, is to somehow reward learning. I mean, that's all I'm saying…. Look, I mean if English departments and History departments can assign papers, I mean why can't Math departments and Engineering departments? Why? Math students are bigger cheaters than history students? I don't buy that. That doesn't make sense to me. (March 17, 2010)

> Michael: If we had an exam where there was one conceptual question—and I screwed up every other question. Just, you know, got every computation wrong. As long as a professor looked at my conceptual answer, and it made sense, I would be happy. (May 13, 2009)

> Michael: If the professors on a one-hour exam actually gave like a serious problem to really test how creative you were, no one would get it right. So, there'd be a few geniuses who could. But see the time constraint is the key, because it shouldn't be about how fast you are. Quick people are rewarded, not deep people. It's about how fast you can cover a lot of surface. (March 17, 2010)

In brief, Michael values deeper conceptual understanding, of the kind that cannot be fully assessed on time-pressured wide-coverage exams.

### II. I'm passionate about learning, but it's a hobby

One might think Michael believes what he believes to justify poor performance on "traditional" problem-solving tasks. *He clings to his ideals,* the argument goes, *because they conveniently privilege a skill at which he excels while disparaging traditional tasks in which he underperforms*. Indeed, Michael has a friend he describes as "brilliant." This friend earned a provisional patent for an engine design but can't seem to keep his grades up.

> Michael: The point is [his engine] was just a work of brilliance. And, he is doing—he has a bad GPA here. Y'know? So he would have trouble finding a job even though he's remarkable at what's going on. (May 13, 2009)

It's tempting to argue that Michael's case must be similar to his friend's—Michael *wants* the system to value what he's good at because he's *not* good at what the system values. Demonstrably, such an argument doesn't hold.

Michael's homeworks and exams in his circuit courses clearly show his prowess at *both* conceptual reasoning and traditional problem-solving tasks. Michael himself is keenly aware that for some people, "your beliefs just value what you're good at." But he doesn't accept this psychological insight as a valid refutation of his views.

> Michael: I got straight A's last semester. A lot of people who use this [sense-making] rhetoric try to find excuses for defending their poor GPAs. I'm not one of those students. I think the reason it hasn't affected my GPA is because I view learning as a hobby. So, as with any hobby, you shouldn't let it interfere your GPA. But it is one of my hobbies, and I do enjoy learning, I just—up to the point where I get my grades done. You know what I mean? (March 17, 2010)

In the intervening year since our first interview, Michael began to stress the importance of compartmentalizing "learning" and deep thinking. Now, he treats them collectively as a hobby. Admittedly, it's a hobby he's still very passionate about. But, it's ultimately something he's learned to compromise on when the system demands it.

*III. I'm probably a fringe as far as students go*

The crucial point we wish to make is that Michael sees himself as an outsider, defined in part by what he jokingly calls his "illness," his need to make sense of things. Further, he's painfully aware that his views about learning and knowledge constitute a kind of counterculture in his engineering program. Consequently, he's willing to temper his ideals for the sake of succeeding in the system. As much as he wishes his program were different, he feels he must play by its rules to succeed: "when [engineering firms] look at resumes, one person has this GPA [and] one person has *this* GPA. They throw one out" (May 13, 2009). Nevertheless, the compromises he makes are hard-fought. He resents having to make them, which feeds back into his views of being an outsider in his engineering program.

Michael's sense of marginalization is so strong that he began our most recent interview, unprompted, with the following warning:

> Michael: I'm not sure how interviewing me is useful for your project. Because I was thinking about it and you know, it's like, I dunno if you realize and maybe I did a bad job of explaining it. But, I'm probably like, *a fringe* as far as students go. As far as my views, you know, my ideals. And so, I'm just curious if you're trying to make like a statistical argument. I'm probably hurting your thesis, whatever it is. (March 17, 2010)

Michael's concerns for our research are heartening, and reflect what we believe is a developed sense of trust between interviewer and interviewee. They also embody the core of our argument. Michael's engineering program produced a student who thinks he's biasing our [the researchers'] assessment of the program, simply by virtue of who he is and how he views learning. He sees himself as a fringe element, and wants to be sure we don't confuse him for "typical." He carries strongly-held beliefs about what engineering education and assessment *could* and *should* be, but will grudgingly give them up if they jeopardize his grades. Finally, success in his program and the larger engineering system—as he has come to cede—is about GPA, not understanding.

MICHAEL'S SENSE-MAKING POSITIONS HIM APART FROM PEERS IN CLASSROOM DISCUSSIONS

Michael's epistemological views, and how they differ from those of his classmates, manifest themselves not just during interviews but also during his classroom interactions with his TA and the other students.

*I. Why is it obvious resistors don't store energy?*

During one discussion section, Adam—another student in Michael's class—asks about calculating the energy in a resistor.

> Adam: And what's the energy equation for a resistor? Or is there not one? There isn't one, right? Because it doesn't—
>
> TA: There's no energy stored
>
> Adam: OK

The TA continues solving the problem, until Michael cuts in about ten seconds later.

> Michael: Why is that so obvious? That there's no energy stored in a resistor? Is it just because all the energy dissipates as heat, right away, or? Like...
>
> Angie: There's no field to store it in.
>
> Michael: Heat doesn't...heat doesn't count as— that's what I'm saying. When it heats up it doesn't count as storing energy?
>
> TA: No. It, it dissipates. It's given off. It's not keeping it in.
>
> Adam: That was like, from our first homework.
>
> TA: Y—yeah. Or second. First or second. I dunno, so, yeah.
>
> Adam: There was a question like that on the first homework, so whatever. One of 'em.

In the above episode, Michael takes a different epistemological stance from Adam. Adam focuses on whether it has been authoritatively established "from our first homework" or "whatever," that a resistor stores no energy. Michael, by contrast, wants to understand *why* the heat generated by a resistor "doesn't count as storing energy." Crucially, Michael's stance doesn't lead him to focus on solving the problem at hand, but rather to create and evaluate *intuitively convincing physical explanations* that bridge mathematics (the presence or absence of an equation for stored energy) and circuit behavior. For Adam, it's enough that the result appeared on a previous homework, as the TA corroborates. For Michael, "school precedent" has far less persuasive weight.

*II. I can't even picture how the graph would look*

In another discussion section, Michael's class is analyzing a mesh circuit problem in which it's not obvious that an entire branch of the circuit is ultimately a distraction—no current flows through it. We lack space to present the specific of the problem; but they are not important to our argument, which is about the warrants students use to justify their claims. Here, we also note that Angie, unlike most other student in the class, has taken several upper-level mathematics courses.

> Michael: Maybe I just haven't taken enough math {glances sidelong at Angie}, but it seems like you have two completely different relationships depending on which way the current's going. Because when the current's going, uh, clockwise…no energy is lost. But when current's going counter-clockwise, energy is, uh—because of the resistor. D'you see what I'm saying? So I—I can't even picture how the graph would look.

Another student, S1, points out current can't flow backward through a diode. Michael realizes that this physical restriction means no current flows through the resistive branch at all. But then, Michael wonders why the unused branch appears in the circuit at all, if no current flows through it.

> S1: It sure makes the math a lot easier. I don't think [the professor]'s evil enough to try to make you do something where it goes backwards through a resistor and the waveform is different in one way...

> TA: Yeah, that's good reasoning.

> Adam: {to S1} Our entire semester's riding on your opinion of [the professor]. Can you handle that pressure?

For S1 and Adam, the topic of discussion is whether the professor would consider a particularly difficult-to-analyze circuit branch as fair game on homework and exams. Michael, by contrast, focuses on sense-making. He summarizes his struggle to make *physical* sense of how the circuit dissipates energy when current flows one way but not the other as his inability to generate a *mathematical* object: "I can't even picture what the graph would look like." Michael's talk thus reflects his commitment to both intuitively convincing physical explanation and to external coherence between such reasoning and graphical representations thereof. As a result, he's less interested in justifications that hinge on what the professor would do, and he does not particulate in that part of the discussion.

## SUMMARY AND DISCUSSION

The engineering education community is taking powerful and important steps to retain students, particularly among demographically underrepresented groups [1],[11],[12],[14],[16]. But, we need to think more broadly about what, in addition to demographic diversity, we're trying to retain. Just as some students see engineering as an intersection of intellectual praxis and social responsibility [28], so too should we begin to think of retention as the intersection of a social *and* intellectual endeavor.

Michael, we argue, should spur discussions about the kinds of disciplinary practices we value and their relationship to the individuals who enact them. We've shown that Michael's identity as a sense-maker embodies a productive, powerful attitude toward engineering that he perceives to be both atypical and undervalued in his courses. If sense-making suffuses successful professional engineering [5],[6],[8],[19]—and we contend it does—then we must ask why Michael feels his views are so out of place in his program.

The suggestion we offer is one of culture. Michael's engineering program, in his view, makes it sensible to sidestep mathematical sense-making because computation and traditional problem-solving are the lowest-hanging fruit for student success. We note the problem "is that the same behavior that is sensible in one context (schooling as an institution) may violate the protocols of sense making in another," in this case the culture of practicing engineers [9]. We agree with Schoenfeld: It's not that students *aren't* sense-making, it's that they're sensibly playing by the rules of an artificial system that doesn't reflect the profession. Michael's identity as a sense-maker makes him beholden to a different set of protocols—ones he's come to believe are on the margin—because they aren't strongly encouraged or rewarded in his classes. If our goal as educators is to develop and retain students like Michael, then part of our task is to create classroom cultures that value the practices characteristic of "Michaels."

## ACKNOWLEDGMENT

We thank "Michael" for participating in our study and the instructor of his Circuits course for providing us access to students for interviews. We thank Eric Kuo and Michael M. Hull for videotaping discussion sections. We thank David Hammer, Eric Kuo, Michael M. Hull, and Indigo Esmonde for productive discussions. Finally, we thank Jennifer Richards for her encouragement, moral support, and uncanny eye for detail. This work was supported in part by NSF EEC-0835880 and NSF DRL-0733613. The opinions are of the authors only.

## REFERENCES

[1] N.L. Fortenberry, J.F. Sullivan, P.N. Jordan, and D.W. Knight, "RETENTION: Engineering Education Research Aids Instruction," *Science*, vol. 317, Aug. 2007, pp. 1175-1176.

[2] D. Hammer, "Epistemological Beliefs in Introductory Physics,"

[3] D. Hammer and A. Elby, "On the Form of a Personal Epistemology," *Personal Epistemology: The Psychology of Beliefs About Knowledge and Knowing*, B.K. Hofer and P.R. Pintrich, Eds., Mahwah, N.J: L. Erlbaum Associates, 2002, pp. 169-190.
[4] J.P. Gee, "Identity as an Analytic Lens for Research in Education," *Review of Research in Education*, vol. 25, 2000, pp. 99-125.
[5] R. Stevens, K. O'Connor, and L. Garrison, "Engineering student identities in the navigation of the undergraduate curriculum," *Proceedings of the 2005 American Society for Engineering Education Annual Conference & Exposition*, 2005.
[6] R. Stevens, K. O'Connor, L. Garrison, A. Jocuns, and D.M. Amos, "Becoming an Engineer: Toward a Three Dimensional View of Engineering Learning.," *Journal of Engineering Education*, vol. 97, Jul. 2008, pp. 355-368.
[7] B.L. Sherin, "How Students Understand Physics Equations," *Cognition and Instruction*, vol. 19, 2001, pp. 479-541.
[8] J. Gainsburg, "The Mathematical Modeling of Structural Engineers.," *Mathematical Thinking & Learning*, vol. 8, Jan. 2006, pp. 3-36.
[9] A.H. Schoenfeld, "On mathematics as sense-making: An informal attack on the unfortunate divorce of formal and informal mathematics," *Informal reasoning and education*, 1991, pp. 311–343.
[10] A.H. Schoenfeld, "Learning to Think Mathematically: Problem Solving, Metacognition, and Sense-Making in Mathematics," *Handbook for research on mathematics teaching and learning*, D. Grouws, Ed., New York: MacMillan, 1992, pp. 334-370.
[11] J.E. Froyd and M.W. Ohland, "Integrated Engineering Curricula.," *Journal of Engineering Education*, vol. 94, Jan. 2005, pp. 147-164.
[12] C.L. Owens and N.L. Fortenberry, "A transformation model of engineering education.," *European Journal of Engineering Education*, vol. 32, 2007, pp. 429-440.
[13] W. Pearson and A. Fechter, Eds., *Who Will Do Science?: Educating the Next Generation*, Baltimore: Johns Hopkins University Press, 1994.
[14] J. Clark Blickenstaff, "Women and science careers: leaky pipeline or gender filter?," *Gender and Education*, vol. 17, 2005, pp. 369-386.
[15] R. Varma and H. Hahn, "Gender and the pipeline metaphor in computing.," *European Journal of Engineering Education*, vol. 33, Mar. 2008, pp. 3-11.
[16] E. Seymour and N.M. Hewitt, *Talking About Leaving: Why Undergraduates Leave the Sciences*, Boulder, Colo: Westview Press, 1997.
[17] M. Yerushalmy, "Designing Representations: Reasoning about Functions of Two Variables," *Journal for Research in Mathematics Education*, vol. 28, Jul. 1997, pp. 431-466.
[18] J. Boaler and J.G. Greeno, "Identity, agency, and knowing in mathematical worlds," *Multiple Perspectives on Mathematics Teaching and Learning*, J. Boaler, Ed., Westport, CT: Ablex Pub, 2000, pp. 171-200.
[19] R.P. Hall and R. Stevens, "Making space: a comparison of mathematical work in school and professional design practices," *The Cultures of Computing*, S.L. Star, Ed., Oxford, UK: Blackwell Publisher, 1995, pp. 118-145.
[20] S. Tobias, *They're Not Dumb, They're Different: Stalking the Second Tier*, Tucson, Ariz. (6840 E. Broadway Blvd., Tucson 85710-2815): Research Corp, 1990.
[21] A.M. Atkin, R. Green, and L. McLaughlin, "Patching the Leaky Pipeline: Keeping First-Year College Women Interested in Science," *Journal of College Science Teaching*, vol. 32, Oct. 2002, pp. 102-108.
[22] J.A. Maxwell, *Qualitative Research Design: An Interactive Approach*, Thousand Oaks, CA: Sage Publications, 2005.
[23] E.G. Guba and Y.S. Lincoln, "Paradigmatic controversies, contradictions, and emerging confluences," *The SAGE Handbook of Qualitative Research*, N.K. Denzin and Y.S. Lincoln, Eds., Thousand Oaks: Sage Publications, 2005, pp. 191-215.
[24] R. Jessor, A. Colby, and R.A. Shweder, Eds., *Ethnography and human development: Context and meaning in social inquiry*, University of Chicago Press, 1996.
[25] M.B. Miles and A.M. Huberman, *Qualitative Data Analysis: A Sourcebook of New Methods*, Beverly Hills: Sage Publications, 1984.
[26] L. Richardson, *Fields of Play: Constructing an Academic Life*, New Brunswick, N.J: Rutgers University Press, 1997.
[27] A.H. Schoenfeld, "When Good Teaching Leads to Bad Results: The Disasters of 'Well-Taught' Mathematics Courses," *Educational Psychologist*, vol. 23, 1988, pp. 145-166.
[28] L.S. Shulman, "If Not Now, When? The Timeliness of Scholarship of the Education of Engineers.," *Journal of Engineering Education*, vol. 94, Jan. 2005, pp. 11-12.

AUTHOR INFORMATION

**Brian A. Danielak**, Doctoral Student in Curriculum and Instruction specializing in Science Education Research, University of Maryland, briandk@umd.edu

**Ayush Gupta**, Research Associate in Physics Education Research, Department of Physics, University of Maryland, ayush@umd.edu

**Andrew Elby**, Assistant Research Scientist in the Departments of Physics and Curriculum and Instruction, University of Maryland, elby@umd.edu